\documentclass{cimento}
\usepackage{amsmath}
\usepackage{amssymb}
\usepackage{slashed}

\usepackage{graphicx} 

\title{Tensor glueball decay into nucleon-antinucleon}
\author{A.~Vereijken \from{ins:x}
}
\instlist{\inst{ins:x}Institute of Physics, Jan Kochanowski University, ul. Uniwersytecka 7, 25-406, Kielce, Poland
}

\begin{document}

\maketitle

\begin{abstract}
 In the context of a chiral hadronic model, we compute the decay ratio of a tensor glueball decaying into a nucleon and antinucleon compared to the decay into 2 pions. Tensor meson dominance is assumed to also hold for the tensor glueball in order to relate the coupling constants of the different decay channels. We find that the decay width to nucleons is slightly larger than the decay width to pions, but still in the same order of magnitude.
\end{abstract}

\section{Introduction}
Glueballs, bound states made of only gluons, are one of the oldest predictions of Quantum Chromodynamics (QCD) \cite{Gross:2022hyw}. Various theoretical (e.g. \cite{Giacosa:2005bw,Brunner:2015oqa,Athenodorou:2020ani}) and experimental \cite{Klempt:2022qjf} works have made progress, yet their experimental status is not resolved \cite{Mathieu:2008me,Crede:2008vw,Llanes-Estrada:2021evz,Chen:2022asf}. Different theoretical methods agree on the mass hierarchy of the lowest lying glueball states, with the scalar $(J^{PC}=0^{++})$ being the lightest and the tensor $(J^{PC}=2^{++})$ the second lightest glueball. In this work, in the context of a chiral model described in \cite{Vereijken:2023jor,Jafarzade:2022uqo}, we will calculate the decay of the tensor glueball into a nucleon-antinucleon pair, based on arguments used in tensor mesons dominance models. Different glueballs have been studied before in hadronic models, such as the scalar glueball in \cite{Janowski:2014ppa} and the pseudoscalar glueball in \cite{Eshraim:2012jv}.

\section{Decay Amplitude}
The decays of the tensor glueball were studied in the extended Linear Sigma Model in \cite{Vereijken:2023jor}, where the $\rho\rho$ and $K^{*}\bar{K}^{*}$ channels were found as the dominant decays. Here, we extend that model by coupling the tensor glueball to nucleons with the following interaction term \cite{Wang:2014yza,Yu:2011zu}
\begin{equation}
    \mathcal{L}_{GNN} =\frac{g_{NN}}{m} G_{\mu\nu} \Bar{N}(\gamma^{\mu} \overset{\leftrightarrow}{\partial^\nu} + \gamma^{\nu} \overset{\leftrightarrow}{\partial^\mu}) N,
\end{equation}
where $G_{\mu\nu}$ is the tensor glueball, $\Bar{N},N$ is the (anti-)nucleon field containing the proton and neutron, i.e. $N^{T} = (p,n)$, $m$ is the nucleon mass, and $\overset{\leftrightarrow}{\partial^\mu} = \overset{\rightarrow}{\partial^\mu} - \overset{\leftarrow}{\partial^\mu}$.
Writing it out explicitly and using the symmetry of $G_{\mu\nu}$ we have the Feynman diagram in figure \ref{feynman} and associated matrix element:
\begin{figure}
\includegraphics{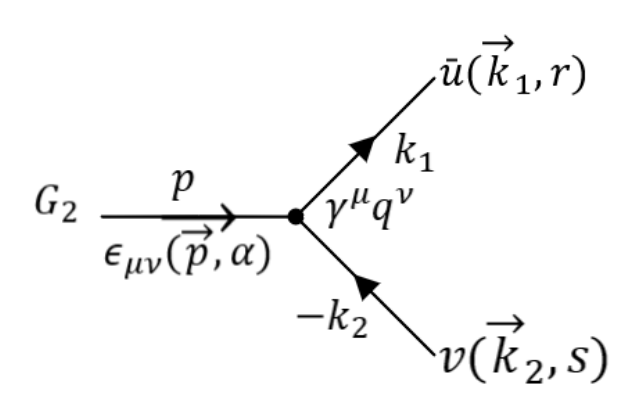}     
\caption{Feynman diagram for tensor glueball decay}
\label{feynman}
\end{figure}

\begin{equation}
    \mathcal{M}(\alpha,r,s) = 2 \frac{g_{NN}}{m} \epsilon_{\mu\nu}(\vec{p},\alpha) \Bar{u}(\vec{k}_{1},r)\gamma^{\mu}q^{\nu}v(\vec{k}_{2},s),
\end{equation}
with $\epsilon_{\mu\nu}(p,\alpha)$ the spin 2 polarization tensors, $\Bar{u}(\Vec{k}_{1},r), v(\vec{k}_{2},s)$ Dirac spinors, and $q = k_{1}-k_{2}$ is the difference of outgoing momenta. Using the Casimir trick, the spin-averaged modulus squared matrix element is
\begin{align}
    \Bar{|\mathcal{M}|}^{2} = \frac{4}{5} \left(\frac{g_{NN}}{m}\right)^2\sum_{\alpha}\epsilon_{\mu\nu}(p,\alpha)\epsilon_{\mu'\nu'}(p,\alpha)q^{\nu}q^{\nu'} \text{Tr}\left[\gamma^{\mu}(\slashed{k}_{2}-m)\gamma^{\mu'}(\slashed{k}_{1}+m) \right]. 
\end{align}
The polarization tensors fulfill the completeness relation \cite{Koenigstein:2015asa}
\begin{equation}
    \sum_{\alpha}\epsilon_{\mu\nu}(p,\alpha)\epsilon_{\mu'\nu'}(p,\alpha) = \frac{1}{2}(A_{\mu\mu'}A_{\nu\nu'}+A_{\mu\nu'}A_{\mu'\nu})-\frac{1}{3}A_{\mu\nu}A_{\mu'\nu'},
\end{equation}
with the tensor $A_{\mu\nu}$ defined as
\begin{equation}
    A_{\mu\nu} = g_{\mu\nu} - \frac{p_{\mu}p_{\nu}}{M^2},
\end{equation}
with $M$ being the mass of the tensor glueball, which we take from lattice QCD to be 2369 MeV \cite{Athenodorou:2020ani}. It is useful to note early on that $p\cdot q = (k_{1}+k_{2})\cdot(k_{1}-k_{2}) = k_{1}^2-k_{2}^2 = 0$ since it is the mass difference of the two daughter particles. This simplifies things because any $p_{\nu}, p_{\nu'}$ from the completeness relation automatically gives 0. The amplitude squared then becomes
\begin{align}
 \Bar{|\mathcal{M}|}^{2} = \frac{4}{5} \left(\frac{g_{NN}}{m}\right)^2 &\left[\frac{1}{2}(g_{\mu\mu'}-\frac{p_{\mu}p_{\mu'}}{M^2})g_{\nu\nu'} + \frac{1}{2}g_{\mu\nu'}g_{\mu'\nu} - \frac{1}{3}g_{\mu\nu}g_{\mu'\nu'} \right] q^{\nu}q^{\nu'} 
 \\ &\left(k_{2}{}_{\alpha}k_{1}{}_{\beta} \text{Tr}[\gamma^{\mu}\gamma^{\alpha}\gamma^{\mu'}\gamma^{\beta}] - m^{2} \text{Tr}[\gamma^{\mu}\gamma^{\mu'}] \nonumber
 \right)
\end{align}
Using well-known trace identities for gamma matrices and contracting with $q$ we find
\begin{align}
    \Bar{|\mathcal{M}|}^{2} = &\frac{16}{5} \left(\frac{g_{NN}}{m}\right)^2\left[\frac{1}{2}(g_{\mu\mu'}-\frac{p_{\mu}p_{\mu'}}{M^2})q^2 + \frac{1}{6}q_{\nu}q_{\nu'} \right] 
    \\ &\left(k_{1}^{\mu}k_{2}^{\mu'} +k_{1}^{\mu'}k_{2}^{\mu} - (k_{1}\cdot k_{2})g^{\mu \mu'} - m^2 g^{\mu\mu'}\right), \nonumber
\end{align}
which simplifies to
\begin{align}
    \Bar{|\mathcal{M}|}^{2} = \frac{16}{5} \left(\frac{g_{NN}}{m}\right)^2\Bigg[&\frac{q^2}{2}\left(-(k_{1}\cdot k_{2}) -3m^2 -2 \frac{(p\cdot k_{1})(p\cdot k_{2})}{M^2}\right) 
    \\
        &+\frac{1}{6}\Big( 2(k_{1}\cdot q)(k_{2}\cdot q) -(k_{1}\cdot k_{2})q^2 -m^2 q^2 \Big) \Bigg] \nonumber
\end{align}
Evaluating all dot products in the rest-frame of the tensor glueball the amplitude takes the form
\begin{align}
    \frac{8}{15} \left(\frac{g_{NN}}{m}\right)^2 \left(3 M^4 -4m^2M^2 - 32m^4 \right).
\end{align}
The decay width for a boson decaying into two nucleons is given by \cite{Giacosa:2012hd}
\begin{equation}
    \Gamma_{GN\Bar{N}} = 2 \frac{\sqrt{\frac{M^2}{4}-m^2}}{8\pi M^2}   \Bar{|\mathcal{M}|}^{2} = \frac{1}{30} \left(\frac{2g_{NN}}{m}\right)^2 \frac{\sqrt{\frac{M^2}{4}-m^2}}{\pi M^2} \left(3 M^4 -4m^2M^2 - 32m^4 \right),
\end{equation}
where the factor 2 counts the $p\bar{p}$ and $n\bar{n}$ modes.
\section{Tensor Meson Dominance and decay ratio}
We cannot compute the decay width without knowing the value of $g_{NN}$. Although this coupling constant is not known experimentally, assuming tensor meson dominance \cite{Yu:2011zu} one has certain relations between couplings of different channels for the tensor meson decays. We will assume these for the tensor glueball as well. The Lagrangian for the decay of the tensor glueball to 2 pions is of the form \cite{Vereijken:2023jor, Jafarzade:2022uqo}
\begin{equation}
    \mathcal{L}_{G\pi\pi} = \frac{g_{\pi\pi}}{M} G_{\mu\nu} \partial^{\mu}\vec{\pi} \partial^{\nu}\vec{\pi}, 
\end{equation}
where $\vec{\pi} = (\pi^{1},\pi^{2},\pi^{3})$ refers to the isospin triplet. This Lagrangian leads to decay width of the form
\begin{equation}
    \Gamma_{G\pi\pi} = 6 \left(\frac{g_{\pi\pi}}{M} \right)^2 \frac{\left(\frac{M^2}{4}-m^2\right){}^{5/2} }{60 \pi  M^2},
\end{equation}
where the factor 6 counts the isospin and identical particle factors. 
Tensor meson dominance (TMD) states that the dominant contribution to the hadron energy momentum tensor $\Theta^{\mu\nu}$ is the tensor mesons $T^{\mu\nu}$. Then, assuming tensor meson dominance leads to the following identity \cite{Yu:2011zu}
\begin{equation}
    \frac{2 g_{NN}}{m} = \frac{g_{\pi\pi}}{M},
\end{equation}
which we will assume is also a valid approximation for the tensor glueball. This allows us to calculate the decay ratio of $G \to N\Bar{N}/G \to \pi\pi$ as
\begin{equation}
    \frac{\Gamma_{GN\Bar{N}}}{\Gamma_{G\pi\pi}} \approx 5.3.
\end{equation}
The decay ratio is large enough to be a relevant factor in the search for a tensor glueball. In comparison to a chiral hadronic model \cite{Vereijken:2023jor} or a holographic model \cite{Hechenberger:2023ljn}, the decay width is lower than the 2-vector channel widths, but larger than the other channels. For example compared to the dominant $\rho\rho$ channel found in 
 \cite{Vereijken:2023jor}, the ratio is $\Gamma_{G\rho\rho}/\Gamma_{GN\bar{N}} \approx 9.6$. The applicability of tensor meson dominance to the tensor glueball is not completely clear, so it is at best an approximate result. However, as an order of magnitude estimation, the outcome can be useful.

\section{Conclusion}
In this note we have computed the tensor glueball decay ratio of the nucleon-antinucleon and 2 pion channels. A speculative assumption of tensor meson dominance relations applying also to the tensor glueball has been made. The approximate result we obtained for the decay into nucleon-antinucleon is larger than the 2 pion ratio but the ratio is still of order 1. Compared to previous works it is not the largest decay channel found. Nevertheless it could still be a fruitful process to investigate in glueball searches.

\newpage
\acknowledgments
The author acknowledges Francesco Giacosa and Shahriyar Jafarzade for useful discussions. The author also acknowledges financial support from the Polish National Science Centre (NCN) via the OPUS project 2019/33/B/ST2/00613.

\end{document}